\documentclass[a4paper]{jpconf}
\usepackage{graphicx}
\begin{document}
\title{The interplay between grand unified and flavour symmetries in a Pati-Salam $\times$ $S_4$ model}

\author{Reinier de Adelhart Toorop}

\address{Nikhef Theory Group, Sciencepark 105, 1098 XG Amsterdam, The Netherlands}

\ead{reintoorop@nikhef.nl}

\begin{abstract}
Both discrete flavour symmetries and Grand Unified symmetries explain apparent structures in the mass sector of the Standard Model. A model that combines both symmetries is therefore very appealing. We construct a model with the $S_4$ flavour symmetry and the Pati-Salam unification. We show that this model can indeed explain many observable relations between the masses of the quarks and leptons and that it is predictive in the neutrino sector. However, the combination of the two symmetries leads to new complications in the Higgs sector and in the running of the renormalisation group equations.
\end{abstract}

\section{Introduction}
\label{sec:intro}
The Standard Model (SM) of particle physics has 19 free parameters. 13 of these are related to the mass sector as the masses and mixing angles and phases of the quarks and leptons. If righthanded neutrinos are added, this number even grows to 22 out of 28.

When the masses and mixing angles are studied in detail, patterns seem to emerge. Some of these can be explained by assuming a discrete flavour symmetry at an high energy scale, others by assuming the symmetry of a grand unified theory (GUT). In section \ref{sec:patterns}, we introduce these patterns and in section \ref{sec:syms} we show how they are related to new symmetries. In section \ref{sec:model}, we present a concrete implementation of this idea in a model \cite{Toorop:2010yh} with an $S_4$ family symmetry and the Pati-Salam (PS) unificartion symmetry. This model has a positive and a negative message (``a beauty and a beast''). The positive point (section \ref{sec:beauty}) is that it is indeed possible to build a model that combines discrete and GUT symmetries and explains many observables in the mass sector. The negative point (section \ref{sec:beast}) is that there are unexpected effects in the combination of the two symmetries. This leads to a very enlarged Higgs sector and to a strong effect on the renormalisation group (RG) behaviour. Due to RG effects, a large part of the naively accessible parameter space is excluded. Section \ref{sec:concl} will conclude.

\section{Structure in the fermion masses}
\label{sec:patterns}
\begin{figure}[h]
\includegraphics[width=22pc]{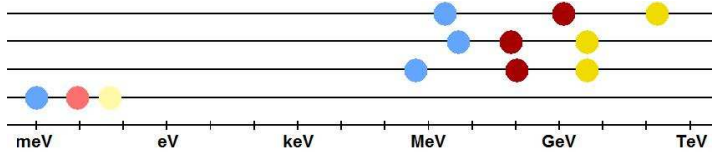}
\hspace{2pc}
\begin{minipage}[b]{14pc}\caption{\label{fig1}The masses of the SM fermions on a logarithmic scale. From first to last line: uptype quarks, downtype quarks, charged leptons and neutrinos (here pictured in the normal hierarchy)}
\end{minipage}
\end{figure}

The mass sector of the Standard Model seems to contain a lot of structure. Firstly the mass gaps between down and strange and between strange and bottom quarks are more or less of the same logarithmic size and of order $\lambda^2$, with $\lambda \sim 0.2$ the Cabibbo angle. The same holds for the charged leptons, while for the up-type quarks, the gaps are somewhat larger, being more or less of order $\lambda^4$. See also figure \ref{fig1}.

\begin{figure}[h]
\begin{center}
\begin{minipage}{14pc}
\includegraphics[width=4pc]{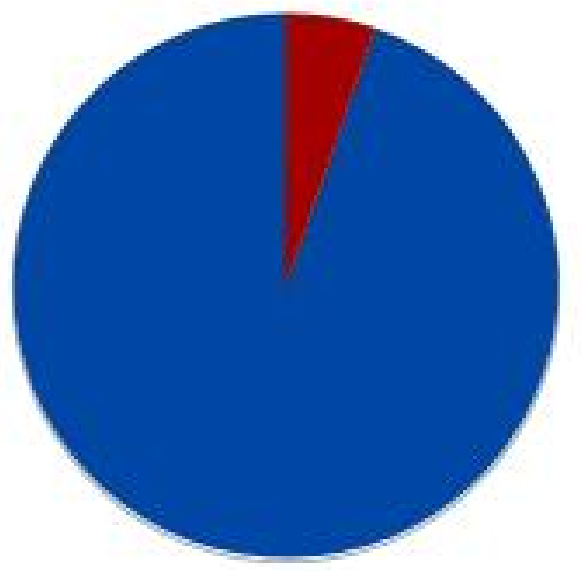}
\includegraphics[width=4pc]{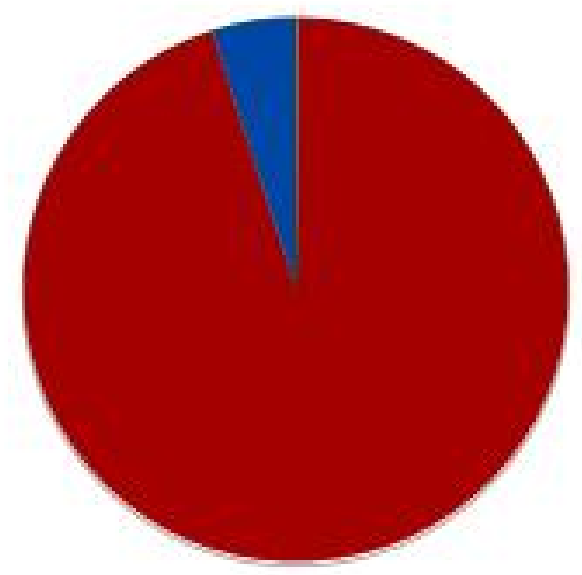}
\includegraphics[width=4pc]{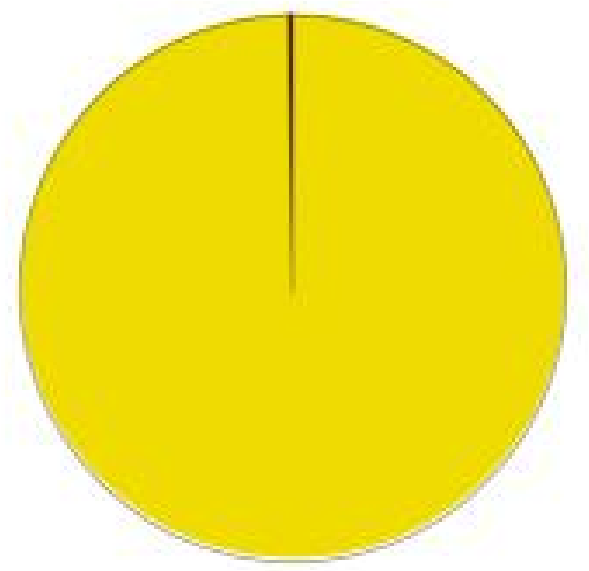}
\caption{\label{fig2}Pie charts showing the flavor content of the three quark mass eigenstates.}
\end{minipage}\hspace{4pc}%
\begin{minipage}{14pc}
\includegraphics[width=4pc]{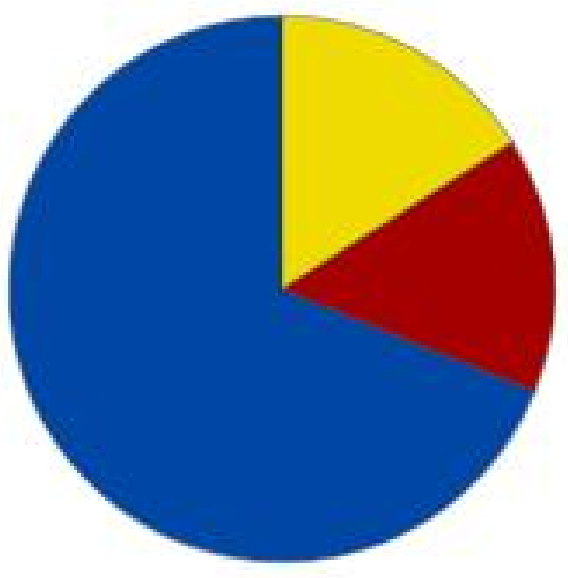}
\includegraphics[width=4pc]{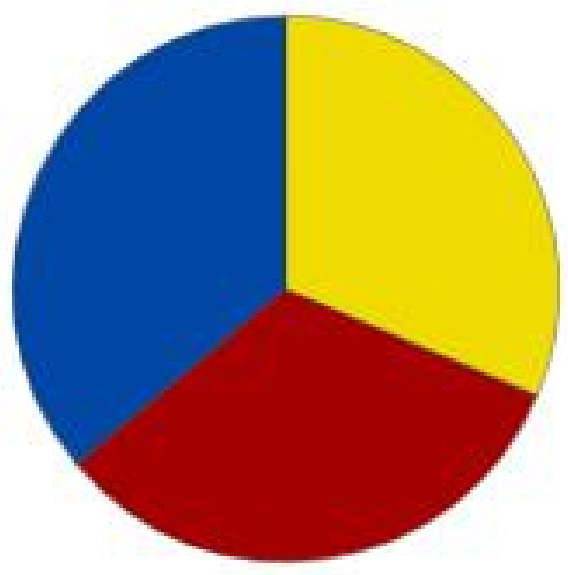}
\includegraphics[width=4pc]{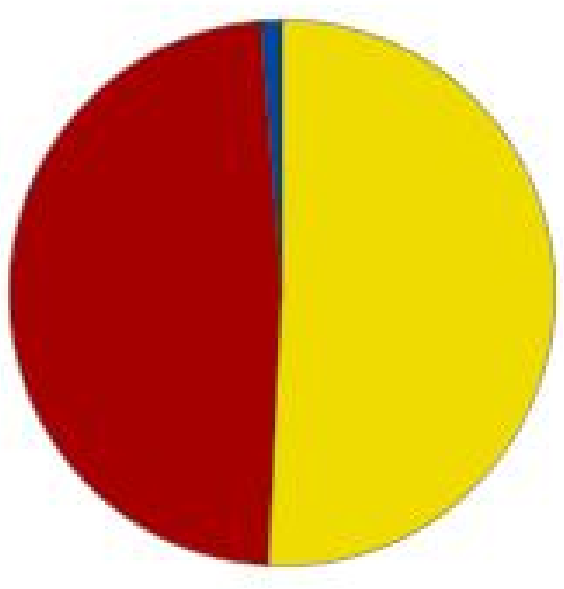}
\caption{\label{fig3}Pie charts showing the flavor content of the three neutrino mass eigenstates.}
\end{minipage}
\end{center}
\end{figure}

A second place where patterns show up, is in the mixing between flavour and mass eigenstates. This mixing is relatively small in the quark sector, where there is one moderate angle ($\theta_{12}^q$, the Cabibbo angle), as well as two small angles ($\theta_{23}^q \sim \lambda^2$ and $\theta_{13}^q \sim \lambda^3$). In the lepton sector, the mixing is much larger and seems to show a particular structure. One neutrino mass eigenstate is an almost equal mixture of all three flavour eigenstates, while another mass eigenstate is a mixture one of two eigenstates (the so called tribimaximal mixing pattern \cite{Harrison:2002er}). See also figures \ref{fig2} and \ref{fig3}. Furthermore, we notice that both in the 12 and in the 23 sector the neutrino and the quark mixing angles add to a maximal angle (45 degrees).

\section{Flavour symmetries and GUTs}
\label{sec:syms}
Flavour symmetries (for an introduction, see e.g. \cite{Altarelli:2006ri}) are symmetries that work between the three family copies of a certain particle type (``a horizontal symmetry''), which is to be contrasted with the usual gauge symmetries that work between the different particle types of one family (``a vertical symmetry'').\\
Family symmetries can explain many of the structures mentioned in the previous section. A simple example is the Froggatt-Nielsen mechanism \cite{Froggatt:1978nt}. In this case, some of the particles are charged under an Abelian $U(1)$ charge, that differs between the families. The $U(1)$ charges are such that the original Yukawa Lagrangian is not invariant. This can be solved by introducing a new scalar field, the so-called flavon $\theta$, that has the opposite charge as some of the fermions. Now terms that include some powers  $\theta / \Lambda_{\textrm{cut-off}}$, are again invariant. Here $\Lambda_{\textrm{cut-off}}$ is the cut-off scale of the theory. If the scalar gets a vacuum expectation value (VEV) that is somewhat smaller than the cut-off, the mass gaps of section \ref{sec:patterns} can be generated with only parameters that are ``naturally'' of order 1. Using non-Abelian groups, mixing patterns such as the tribimaximal one can easily be generated. In particular the discrete groups $A_4$ and $S_4$ are often used.

A second reason to consider flavour symmetries is that these can unify the description of the fermions. If the symmetry group has three-dimensional representations (as for instance $A_4$ and $S_4$ do), the fermions can be represented as one representation of a large group instead of as multiple disjunct copies. The same can be accomplished by some grand unified groups. Their name states that they unify the gauge couplings, but they also do so for the fermions. For instance in the Pati-Salam GUT, instead of six representations to describe one family of quarks and leptons (including a righthanded neutrino), one can do with two. We will see that in the Pati-Salam $\times S_4$ model, instead of $3 \times 6 = 18$ representations, one can use only four. See figure 4.

\begin{figure}[!h]
\begin{center}
\begin{minipage}[t]{15 cm}
\begin{center}
\includegraphics[width=9pc]{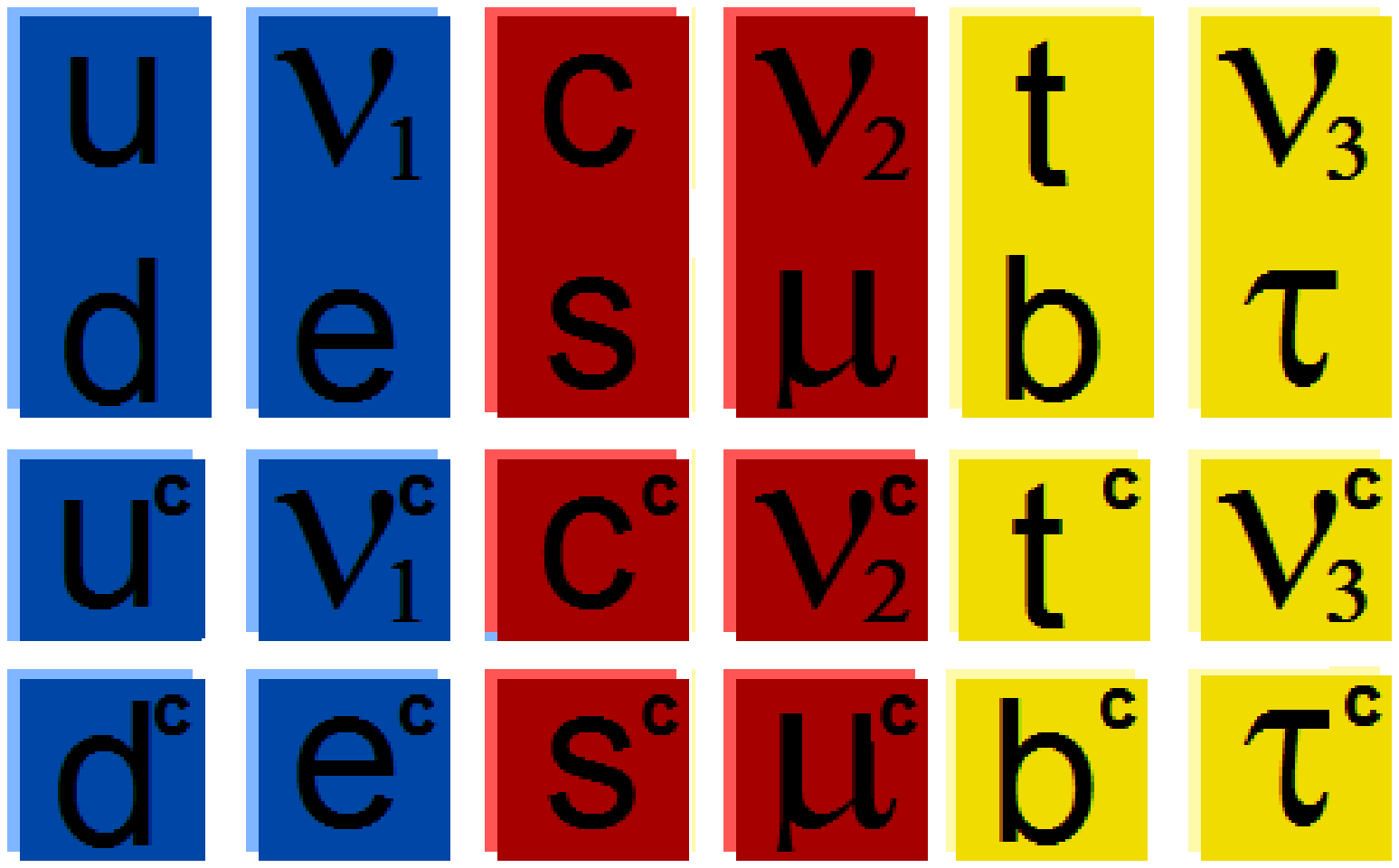}
\qquad
\includegraphics[width=10.9pc]{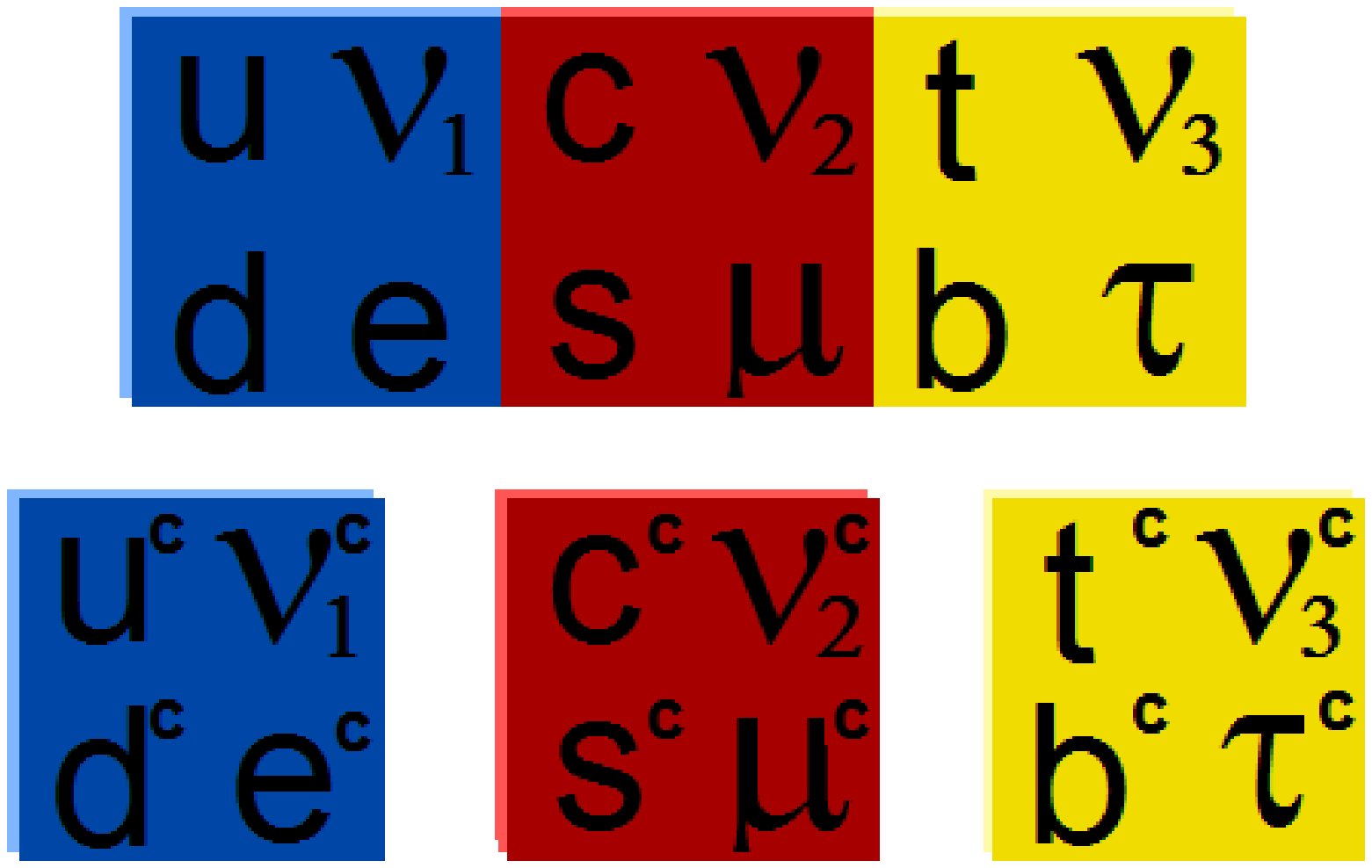}
\end{center}
\end{minipage}
\begin{minipage}[t]{16.5 cm}
\caption{Fermion representations in the Standard Model and in the PS $\times S_4$ model.\label{fig4}}
\end{minipage}
\end{center}
\end{figure}

On the other hand, relations between different particle types of the same generation can often be produced by GUTs. Examples are bottom-tau unification ($m_b = m_\tau$) and the Georgi-Jarlskog relation ($m_\mu = 3 m_s$). We conclude that a model that combines flavour and GUT symmetries is appealing, both on grounds of predictivity and of unification.

\section{The model}
\label{sec:model}
We build an extension of the Standard Model, where the family symmetry group is $S_4$ and the GUT group is Pati-Salam. For more details, see \cite{Toorop:2010yh}. The model uses a supersymmetrical context, although this is not strictly necessary: a non-supersymmetric version of the model with the same main ingredients could be constructed as well.

Technically, the Pati-Salam gauge group $SU(4)_c \times SU(2)_L \times SU(2)_R$ is not a unified group, as even at high energies, there are still three gauge groups: the $SU(4)$ of extended colour and $SU(2)$s for left- and righthanded particles. However, we see that for the fermion representations, there is an unifying aspect, as the lefthanded quark and lepton ($SU(2)_L$)-doublets are combined in one PS representation ({\bf 4}, {\bf 2}, {\bf 1}), as are the four righthanded ones (into ($\mathbf{\bar{4}}$, {\bf 1}, {\bf 2})). For each family, the fermions are now grouped in two representations (that we will denote as $F$ and $f^c$), instead of in six ($Q$, $L$, $u^c$, $d^c$, $l^c$ and $\nu^c$).

The discrete group $S_4$ is the permutation group of four objects and has 24 elements, grouped in five conjugate classes. It has two one-dimensional representations (denoted $1_1$ and $1_2$), one two-dimensional representation ($2$) and two three-dimensional representations ($3_1$ and $3_2$). In particular the fact that there are three-dimensional representations is important, as this allows us to put the lefthanded multiplets of the three families in one representation, thereby providing more unification. We do not put the righthanded multiplets  together in a three dimensional multiplet as well. Instead put them in three separate one-dimensional representations (of type $1_2$, $1_2$ and $1_1$ respectively). This has the advantage that we can charge the families under $U(1)_{FN}$, such that the Froggatt-Nielsen mechanism, alluded to in section \ref{sec:syms} can give suppression to the masses of the first and second generation. The fields are also charged under an extra Abelian $Z_4$ that is used to prevent unwanted couplings in the superpotential. The assignment for the matter fields is summarized in figure \ref{fig4} and table \ref{tab1}.

\begin{table}[!h]
\begin{center}
\caption{Transformation properties of the matter fields.}
\label{tab1}
\begin{tabular}{c c c c c}
\br
 & $F_L$ & $F^c_1$ & $F^c_2$& $F^c_3$ \\
 \mr
  PS & $(4,2,1)$ & $(\overline{4},1,2)$ & $(\overline{4},1,2)$& $(\overline{4},1,2)$  \\
  $S_4$ & $3_1$ & $ 1_2$ & $1_2$ & $1_1$\\
  $U(1)_{FN}$ & 0 & 2 & 2 & 0  \\
  $Z_4 $ & $1$ & $1$ & $i$  & $-i$  \\
  \br
  \end{tabular}
\end{center}
\end{table}

Next to the Froggatt-Nielsen messenger $\theta$, we need four more flavons, three $S_4$ triplets and one singlet. Their properties are given in table \ref{tab2}. The vacuum expectation values mentioned follow from the minimization and the typical magnitude is approximately $\lambda \sim 0.2$ times the cut-off scale of the theory. The VEVs are valid to first order in the number of flavons present in that potential. Subleading terms in this potential have one extra flavon inserted; due to the relative small VEVs, these give relative corrections of order $\lambda$. In fact, these corrections are very well calculable and are crucial in reproducing the observed patterns.

\begin{table}[!h]
\begin{center}
\caption{The flavon field content and their transformation properties under the flavour symmetries. All flavon fields are singlet of the gauge group.}
\label{tab2}
\begin{tabular}{cccccc}
\br
&$\theta$ &  $\varphi$& $\varphi'$& $\chi$ &$\sigma$ \\
\mr
  $S_4$ &$1_1$ & $3_1$ & $3_2$ & $3_1$ &$1_1$ \\
  $U(1)_{FN}$ & -1 &  0 & 0& 0 & 0 \\
  $Z_4$ & 1 & $i$& $i$ &$1$ & $1$\\
 VEV $\propto$ & 1 &
 $\left( \begin{array}{c}
        0 \\
        1 \\
        1 \\
        \end{array}
        \right)$ &
 $\left( \begin{array}{c}
        0 \\
        1 \\
        -1 \\
        \end{array}
        \right)$ &
  $\left( \begin{array}{c}
        0 \\
        0 \\
        1 \\
        \end{array}
        \right)$ &
 $1$  \\
\br
\end{tabular}
\end{center}
\end{table}

Lastly, we will need a number of Higgs fields that break the gauge symmetry and are crucial for obtaining fermion masses. The Higgs spectrum will include at least a field $\phi$ that acts as the electroweak Higgs boson and some Higgs fields that break the PS gauge group to the (supersymmetric) standard model at high energies. The Higgs fields that are relevant for the mass matrices are given in table \ref{tab3}. More details of the Higgs sector are given in section \ref{sec:beast}.

\begin{table}[!h]
\begin{center}
\caption{The Higgs fields  responsible of generating fermion mass matrices and their transformation under the gauge and the Abelian  flavour symmetries. All Higgs fields are singlets under $S_4\times U(1)_{FN}\times U(1)_R$, while they can transform under the $Z_4$ factor.}
\label{tab3}
\begin{tabular}{c c c c c }
\br
 &  ${\phi},{\phi}'$& $\rho$& $\Delta_L$ &$\Delta_R$\\
\mr
  PS & $(1,2,2)$& $(15,2,2)$ & $(\overline{10},3,1)$& $(10,1,3)$ \\

  $Z_4 $ & $1$ & $-1$ & $1$ &$-1$ \\
\br
  \end{tabular}
\end{center}
\end{table}

\subsection{The beauty}
\label{sec:beauty}
As mentioned in the introduction, the model has both a good and a bad side. The good news is that it can explain the patterns mentioned in sections \ref{sec:patterns} and \ref{sec:syms}, as we will show in this section.

At leading order (LO) in the number of flavons, we find mass matrices that can well produce the masses of the second and the third generation, while the first generation is massless. Given that these particles are very light, this is a satisfying first approximation.

The fermion mixing matrices ($V_{CKM}$ and $U_{PMNS}$) however, do not fit the data well: the CKM matrix is the unity matrix, while the PMNS matrix is the so-called bimaximal mixing matrix. The bimaximal mixing matrix has both a maximal atmospheric angle and a maximal solar angle. This result is relatively fine in for the atmospheric angle, where the deviation seems to be small (at most of order $\lambda^2$), while a correction of order $\lambda$ is needed for the solar angle. By definition, we also need a correction of size $\lambda$ to introduce the Cabibbo angle in the CKM matrix. Fortunately, next-to-leading order (NLO) corrections are tailored to be exactly of that order and these can indeed be such that the $12$ angles of in both the quark and lepton sector are corrected, while the $23$ angles only receive next-to-next-to-leading order corrections. The quark-lepton complementarity of section \ref{sec:patterns} is reproduced: the quark and lepton angles add in both cases to a maximal angle. It turns out that the last mixing angle $\theta_{13}$ is relatively large in the neutrino sector (giving an reactor mixing angle of order $\lambda$), while it is indeed tiny $(\mathcal{O} (\lambda^3))$ in the quark sector.

\begin{figure}[h]
\begin{center}
\begin{minipage}{14pc}
\includegraphics[width=4pc]{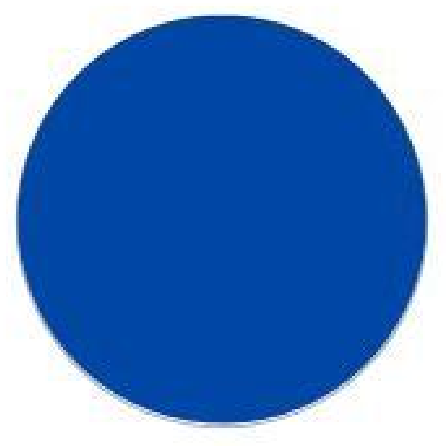}
\includegraphics[width=4pc]{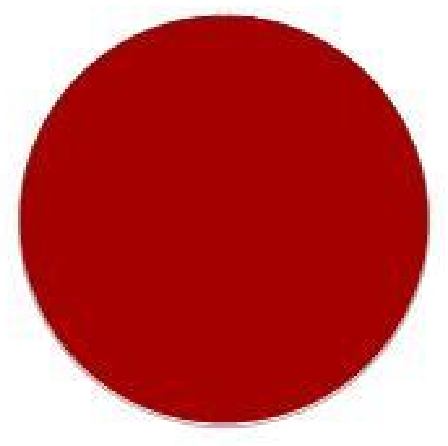}
\includegraphics[width=4pc]{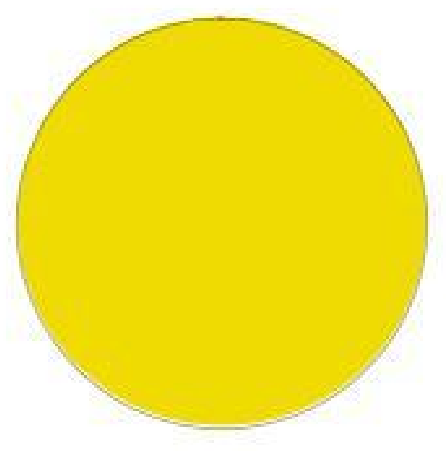} \\
\includegraphics[width=4pc]{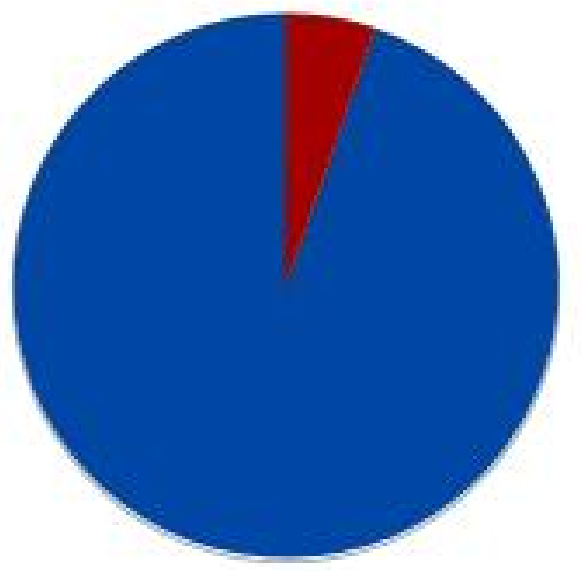}
\includegraphics[width=4pc]{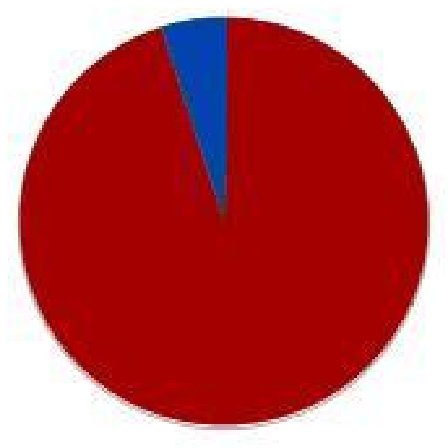}
\includegraphics[width=4pc]{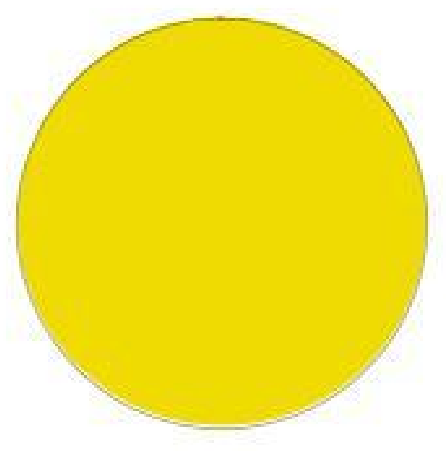}
\caption{\label{fig5}Pie charts showing the flavor content of the three quark mass eigenstates at LO (upper line) and at NLO, when the Cabibbo angle is introduced (lower line). The other two quark mixing angles are generated at even higher order}
\end{minipage}\hspace{4pc}%
\begin{minipage}{14pc}
\includegraphics[width=4pc]{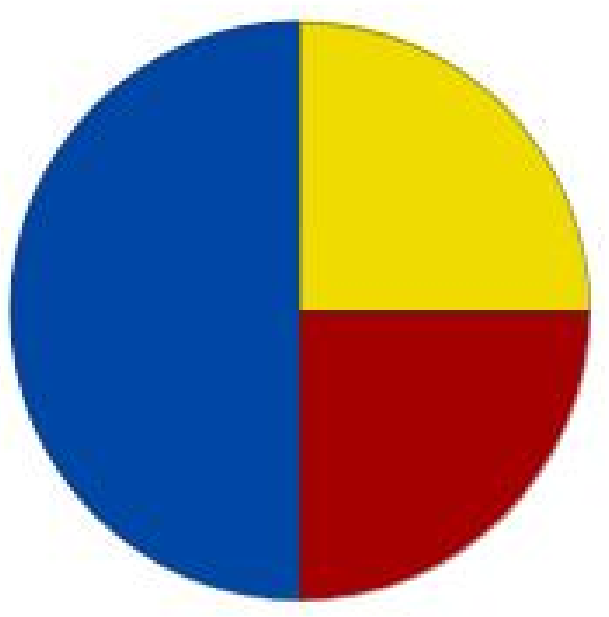}
\includegraphics[width=4pc]{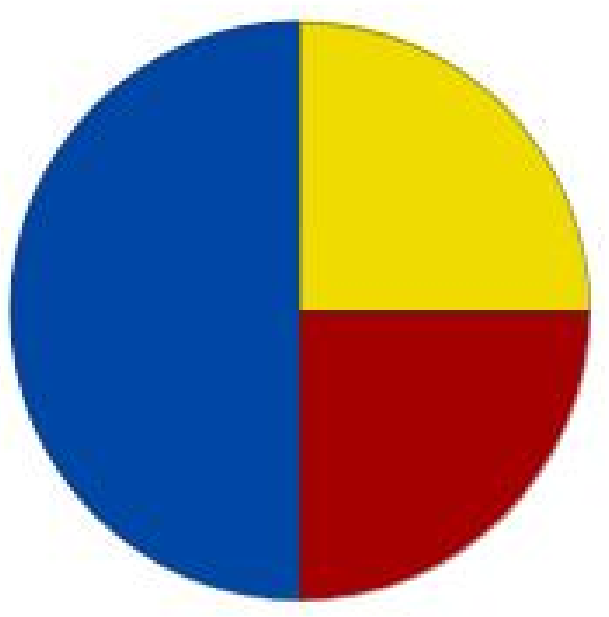}
\includegraphics[width=4pc]{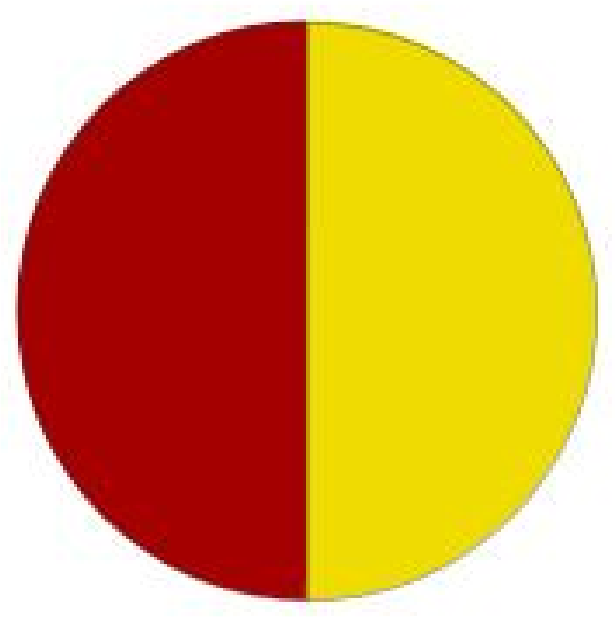}\\
\includegraphics[width=4pc]{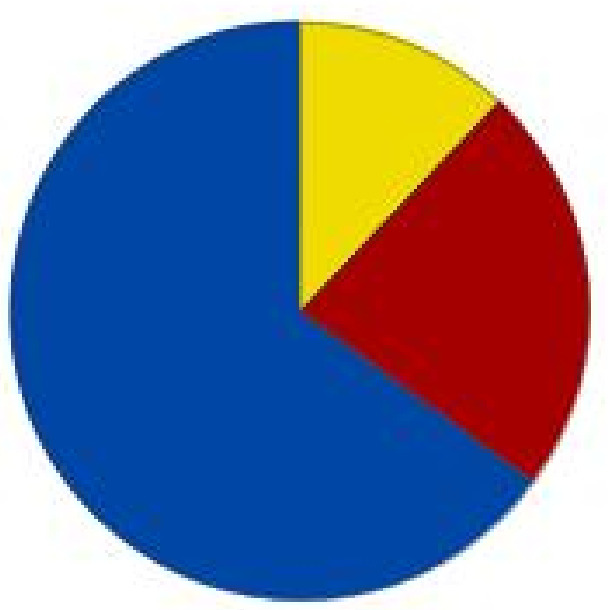}
\includegraphics[width=4pc]{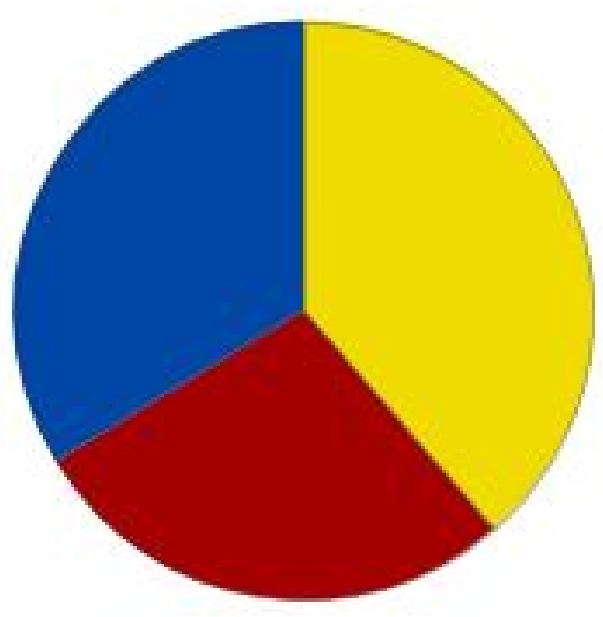}
\includegraphics[width=4pc]{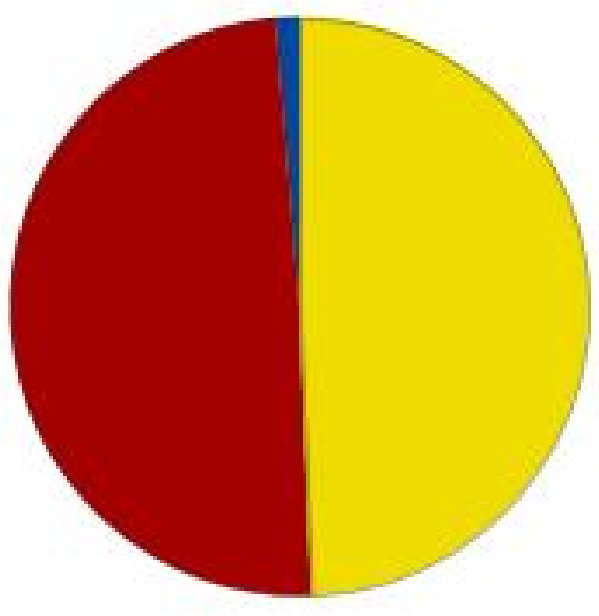}
\caption{\label{fig6}Pie charts showing the bimaximal mixing prediction in the lepton sector at LO (upper line) and a typical NLO result (lower line)}
\end{minipage}
\end{center}
\end{figure}

In a Pati-Salam context furthermore, the matrices of downquarks and charged leptons are connected, as they both particles are in the same Pati-Salam representations. This allows the bottom-tau unification and the Georgi-Jarlskog relations and it helps in connecting the NLO corrections of quark and lepton mixing. The idea of bimaximal mixing with quark complementarity is shown in figures \ref{fig5} and \ref{fig6}.

Apart from explaining quark and lepton mass relations, the model is also predictive in the neutrino sector. Above, it was already mentioned that we predict the third neutrino mixing angle to be relatively large (of the order of the Cabibbo angle). For almost all points in parameter space, the reactor mixing angle should be measureable by next generation experiments such as Daya Bay or Double CHOOZ. A second prediction is that the quasi-degenerate hierarchy is favoured in most of parameter space. Also the normal and inverted hierarchy are possible, albeit only for slightly unnatural values of the parameters. However, we will see that this prediction is severely affected by ``interference'' effects between the two types of symmetries that are analysed in the next section.

\subsection{The beast}
\label{sec:beast}
The positive results of the previous section are somewhat overshadowed by the fact that the combination of the flavour and the GUT symmetry leads to extra effects, which are less appealing. These appear at three points: the Higgs sector and the gauge and Yukawa renormalization group behaviour.

The first point is the Higgs sector, which is very non-minimal in our model. A minimal Pati-Salam model can have only four Higgs fields. These are labeled ``minimal'' in table 4. Other Higgs fields only occur in more extended versions of Pati-Salam models. Examples are the $\Delta_L$ that features in the type II seesaw and $\rho$ that couples to some particles of the second generation and as such generates the Georgi-Jarlskog relation and the stronger suppression of the charm quark. Our construction features even more fields, which is due to the extra $Z_4$ symmetry that forces to double some fields and to the fact that some Higgs fields need to have VEV structures that require the presence of extra fields.

\begin{table}[!h]
\begin{center}
\caption{The Higgs fields of our model. Fields marked ``minimal'' are present in minimal Pati-Salam models; fields marked ``extended'' only in more extended models and fields marked ``new'' are only used in our construction.}
\label{tab4}
\begin{tabular}{c c c c c c c}
\br
& $A$ & $B$ & $\Delta_R$ & $\bar{\Delta}_R$ & $\phi$ & $\phi'$ \\
\mr
PS & (15,1,1) & (15,1,1) & (10,1,3) & ($\bar{10}$,1,3) & (1,2,2) & (1,2,2) \\
$Z_4$ & 1 & -1 & -1 & -1 & 1 & 1 \\
& minimal & new & minimal & minimal & minimal & new \\
\br
& $\rho$ & $\Delta_L$ & $\bar{\Delta}_L$ & $\Sigma$ & $\Sigma'$ & $\xi$ \\
\mr
PS & (15,2,2) & ($\bar{10}$,3,1) & (10,3,1) & (1,3,3) & (1,3,3) & (1,1,1) \\
$Z_4$ & -1 & 1 & 1 & 1 & -1 & -1 \\
& extended & extended & extended & extended & new & new \\
\br
\end{tabular}
\end{center}
\end{table}

This large Higgs sector has a strong effect on the running of the gauge and Yukawa couplings. In particular in the neutrino sector the effects are large. We see that the atmospheric angle might get corrections that drive it too far away from the maximal angle that is predicted at LO and NLO. The bad news is that this happens in particular when then the neutrinos are in the quasi-degenerate hierarchy and in section \ref{sec:beauty} we concluded that this is the case in most of parameter space. We conclude that the quasi-degenerate hierarchy is excluded, but that the slightly less natural inverse and normal hierarchies are still possible, with now the inverse hierarchy favoured in most of the remaining phase space.

\section{Conclusions}
\label{sec:concl}
We conclude by stating once again that a model that combines flavour symmetries and grand unified symmetries is very appealing. A possible choice is the Pati-Salam $\times$ $S_4$ model described here. We have seen that on first sight, the model can explain many observed patterns in the fermion mass sector. However, on second sight, this is not without a price. The simplicity in the fermion sector (4 instead of 18 representations needed) is partly lost to a much larger
Higgs sector. Also a detailed analysis of the running of the gauge and Yukawa couplings showed that much of the naiely preferred phase space is in fact no longer viable.

\end{document}